\documentclass[aps,prl,twocolumn,showpacs,amsmath,amssymb,superscriptaddress]{revtex4}
\usepackage{graphicx}
\usepackage{amssymb}
\usepackage{natbib}
\usepackage{color}

\newcommand{\beq}{\begin{eqnarray}}
\newcommand{\eeq}{\end{eqnarray}}

\begin{document}

\title{Orbital selective spin excitations and their impact on superconductivity of LiFe$_{1-x}$Co$_x$As}
\author{Yu Li}
\affiliation{Department of Physics and Astronomy,
Rice University, Houston, Texas 77005, USA}

\author{Zhiping Yin}
\email{yinzhiping@bnu.edu.cn}
\affiliation{Center for Advanced Quantum Studies and Department of Physics, Beijing Normal University, Beijing 100875, China}
\affiliation{Department of Physics, Rutgers University, Piscataway, New Jersey 08854, USA}

\author{Xiancheng Wang}
\affiliation{Beijing National Laboratory for Condensed Matter
Physics, Institute of Physics, Chinese Academy of Sciences, Beijing
100190, China}

\author{David W. Tam}
\affiliation{Department of Physics and Astronomy,
Rice University, Houston, Texas 77005, USA}

\author{D. L. Abernathy}
\affiliation{Quantum Condensed Matter Division, Oak Ridge National Laboratory, Oak Ridge, Tennessee 37831, USA}

\author{A. Podlesnyak}
\affiliation{Quantum Condensed Matter Division, Oak Ridge National Laboratory, Oak Ridge, Tennessee 37831, USA}

\author{Chenglin Zhang}
\affiliation{Department of Physics and Astronomy,
Rice University, Houston, Texas 77005, USA}

\author{Meng Wang}
\affiliation{Department of Physics, University of California, Berkeley, California 94720, USA}

\author{Lingyi Xing}
\affiliation{Beijing National Laboratory for Condensed Matter
Physics, Institute of Physics, Chinese Academy of Sciences, Beijing
100190, China}

\author{Changqing Jin}
\affiliation{Beijing National Laboratory for Condensed Matter
Physics, Institute of Physics, Chinese Academy of Sciences, Beijing
100190, China}

\affiliation{Collaborative Innovation Center of Quantum Matter, Beijing, China}

\author{Kristjan Haule}
\affiliation{Department of Physics, Rutgers University, Piscataway, New Jersey 08854, USA}

\author{Gabriel Kotliar}
\affiliation{Department of Physics, Rutgers University, Piscataway, New Jersey 08854, USA}

\author{Thomas A. Maier}
\affiliation{Center for Nanophase Materials Sciences and Computer Science and Mathematics Division, Oak Ridge National Laboratory, Oak Ridge, Tennessee 37831, USA}

\author{Pengcheng Dai}
\email{pdai@rice.edu}
\affiliation{Department of Physics and Astronomy,
Rice University, Houston, Texas 77005, USA}
\affiliation{Center for Advanced Quantum Studies and Department of Physics, Beijing Normal University, Beijing 100875, China}

\date{\today}

\begin{abstract}
We use neutron scattering to study spin excitations in single crystals
of LiFe$_{0.88}$Co$_{0.12}$As, which is located near the boundary of the superconducting phase of
LiFe$_{1-x}$Co$_{x}$As and exhibits non-Fermi-liquid behavior indicative of a quantum critical point.
By comparing spin excitations of LiFe$_{0.88}$Co$_{0.12}$As with a combined
density functional theory (DFT) and
dynamical mean field theory (DMFT) calculation, we conclude that wave-vector correlated
low energy spin excitations are mostly from the $d_{xy}$ orbitals, while high-energy spin excitations
arise from the $d_{yz}$ and $d_{xz}$ orbitals. Unlike most iron pnictides, the strong orbital selective spin excitations in LiFeAs family cannot be described by anisotropic Heisenberg Hamiltonian.
While the evolution of low-energy spin excitations of LiFe$_{1-x}$Co$_x$As are consistent with electron-hole Fermi surface nesting
condition for the $d_{xy}$ orbital, the reduced superconductivity in LiFe$_{0.88}$Co$_{0.12}$As
suggests that Fermi surface nesting conditions for the $d_{yz}$ and $d_{xz}$ orbitals are also
important for superconductivity in iron pnictides.
\end{abstract}

\maketitle

Superconductivity in iron pnictides occurs near the vicinity of an antiferromagnetic (AF)
instability \cite{kamihara,cruz,qhunag,Scalapino,pcdai}.
One exception is LiFeAs,
which exhibits superconductivity at $T_c=18$ K without
an AF ordered parent compound \cite{xwang08,Pitcher08,Tapp08}.
Although magnetism is generally believed to play a central role in the superconductivity of iron pnictides \cite{Scalapino,pcdai}, the
unique nature of LiFeAs has raised considerable debates concerning whether magnetism is indeed fundamental to the
superconductivity of iron-based superconductors.
There are two important issues to be addressed. The first is whether magnetism and superconductivity in LiFeAs
can arise from quasiparticle excitations between hole and electron
 nested Fermi surfaces similar to other iron pnictide superconductors \cite{Hirschfeld,Chubukov,YWang13}.
The second concerns the impact of
orbital degrees of freedom on the superconductivity of LiFeAs \cite{TSaito,Saito}.

In most iron pnictides, Fe ions are in a $d^6$ configuration with five same-spin electrons in the $e_g$ and $t_{2g}$ orbitals, and one remaining opposite-spin electron fluctuating among all the $d$ orbitals, due to the large Hund's rule coupling, although there is a considerable (but smaller)
crystal-field splitting between
the $e_g$ and $t_{2g}$ orbitals \cite{Haule_NJP2009,Georges,cclee,kruger,lv,ccchen,valenzeula}.
The $t_{2g}$ electrons near the Fermi level occupy the $d_{xy}$ and degenerate $d_{xz}/d_{yz}$
orbitals. In the undoped state, low-energy spin excitations in LiFeAs are transversely incommensurate from the
AF ordering wave vector of iron pnictides such as BaFe$_2$As$_2$ [Fig 1(a) and Fig. 1(b)] \cite{qhunag}, consistent with nested Fermi surfaces from either the large $d_{xy}$ or small $d_{yz}/d_{xz}$
hole pocket near $\Gamma$ point in reciprocal space
to electron pockets near $M$ points [Fig. 1(c) and Fig. 1(e)] \cite{Qureshi12,MWang12,Qureshi14}.
When Co is doped into LiFeAs to form LiFe$_{1-x}$Co$_{x}$As, superconductivity is gradually suppressed with increasing Co doping and vanishes near $x=0.14$ \cite{MJPitcher10}, and the system becomes paramagnetic
for higher Co-doping levels [Fig. 1(a)] \cite{YMDai}.  From angle resolved photoemission spectroscopy
(ARPES) \cite{ZRYe,HMiao15},
it was found that Co-doping introduces electrons to LiFeAs, reduces the size of the $d_{xy}$ hole Fermi surface,
moves the small $d_{yz}/d_{xz}$
hole pockets below the Fermi surface, and enlarges the electron pockets [Fig. 1(d)].  While the hole-electron
Fermi surface nesting condition is improved for the $d_{xy}$ orbitals near $x=0.12$,
Fermi surface nesting is no longer possible for the $d_{yz}/d_{xz}$ orbitals [Fig. 1(d)].
Since transport,
optical spectroscopy, and nuclear magnetic resonance measurements on LiFe$_{1-x}$Co$_{x}$As
find enhanced low-energy spin fluctuations
near $x=0.12$ with non-Fermi liquid behavior,
these results were taken as evidence that spin fluctuations due to enhanced
Fermi surface nesting can give rise to the observed non-Fermi liquid behavior, but
are not important for superconductivity of LiFeAs \cite{YMDai}.

In this Letter, we present inelastic neutron scattering study
and a combined
density functional theory (DFT) and
dynamical mean field theory (DMFT)
calculation of spin excitations in LiFe$_{0.88}$Co$_{0.12}$As.
While low-energy spin excitations in LiFe$_{0.88}$Co$_{0.12}$As indeed
become commensurate consistent with
improved electron-hole Fermi surface nesting condition for the $d_{xy}$ orbitals [Fig. 1(d)],
the absence of the hole Fermi pockets near the $\Gamma$ point prevents the electron-hole nesting between
the $d_{yz}/d_{xz}$ orbitals. Since our DFT+DMFT calculations suggest
a strongly correlated $d_{xy}$ orbital with much reduced magnetic bandwidth and effective exchange coupling
(Fig. 2, 3, 4), the improved nesting condition in LiFe$_{0.88}$Co$_{0.12}$As,
while sufficient to induce the observed non-Fermi liquid behavior \cite{YMDai}
and increased magnetic excitations near the AF wave vector, is insufficient to cause
superconductivity due to increased incoherent electronic state of the $d_{xy}$ band in
Co-doped LiFeAs \cite{ZRYe}.  Similarly, we find that
spin excitations at higher energies with much steeper dispersion arise mostly
from electron-hole quasiparticle excitations of the $d_{yz}/d_{xz}$ orbitals with much larger magnetic bandwidth and effective exchange coupling compared with NaFeAs (Fig. 4) \cite{CLZ14}. Therefore, spin excitations in the LiFeAs family are highly orbital selective. While spin waves in many iron pnictides can be well described by an anisotropic Heisenberg Hamiltonian \cite{pcdai}, the spin excitations in the LiFeAs family
cannot be satisfactorily explained by such a model.
Our results thus suggest that
the occurrence of
superconductivity in LiFe$_{1-x}$Co$_{x}$As requires Fermi surface nesting of the $d_{xz/yz}$ orbitals.

\begin{figure}[t]
\includegraphics[scale=.48]{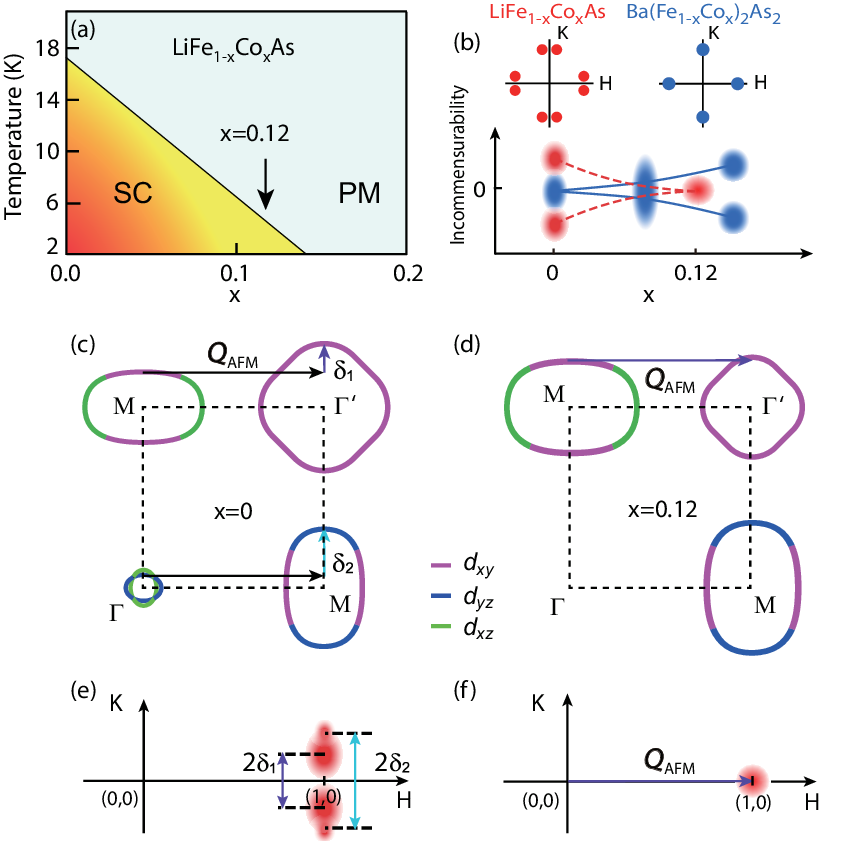}
\caption{(Color online) (a) Electronic phase diagram of LiFe$_{1-x}$Co$_x$As. The superconductivity (SC) is suppressed
by Co-doping and the system is in paramagnetic (PM) phase above $T_c$. The arrow indicates the doping level of $x=0.12$ in our experiment \cite{ZRYe,HMiao15}. (b) Evolution of the low energy spin excitations in reciprocal space
with electron doping for LiFeAs and BaFe$_2$As$_2$. Red spots indicate positions of low energy spin fluctuations in
LiFe$_{1-x}$Co$_x$As and Blue ones are for BaFe$_{2-x}$Ni$_x$As$_2$ \cite{hqluo}. Schematics of the Fermi surfaces for LiFeAs (c) and LiFe$_{0.88}$Co$_{0.12}$As (d) \cite{ZRYe}. Based on ARPES measurements, the mismatched hole and electron Fermi surfaces should result in
the incommensurate spin excitations at $\delta_1$ and $\delta_2$. (e) Positions of
transverse incommensurate spin excitations of LiFeAs at $E=10$ meV seen in the neutron scattering measurements \cite{Qureshi14}. (f) Commensurate spin excitations of LiFe$_{0.88}$Co$_{0.12}$As at $E=10$ meV.}
\end{figure}

Our inelastic neutron scattering measurements on LiFeAs and LiFe$_{0.88}$Co$_{0.12}$As were carried out at the
wide Angular-Range Chopper Spectrometer (ARCS) and Cold Neutron Chopper Spectrometer (CNCS) at Spallation
Neutron Source, Oak Ridge National Laboratory.
Single crystals of
LiFeAs (3.95-g) and LiFe$_{0.88}$Co$_{0.12}$As (7.58-g) are grown using flux method with $^7$Li isotope.
We define the momentum transfer ${\bf Q}$ in three-dimensional reciprocal space in \AA$^{-1}$
as $\textbf{Q}=H\textbf{a}^\ast+K\textbf{b}^\ast+L\textbf{c}^\ast$, where $H$, $K$, and $L$ are Miller indices and
${\bf a}^\ast=\hat{{\bf a}}2\pi/a$, ${\bf b}^\ast=\hat{{\bf b}}2\pi/b$,
${\bf c}^\ast=\hat{{\bf c}}2\pi/c$ with  $a= b\approx 5.316$ \AA,
and $c=6.306$ \AA\ for both samples.
In this notation, the AF Bragg peaks for magnetically ordered compound NaFeAs
should occur at ${\bf Q}_{AF}=(\pm 1,0,L)$ ($L=0.5,1.5,\cdots$)
positions in reciprocal space [Fig. 1(e) and 1(f)] \cite{SLLi}.  Samples are co-aligned
in the $[H,0,L]$ scattering plane with mosaic less than
3$^\circ$ and incident beam ($E_i=20,35,80,250,450$ meV)
parallel to the $c$-axis of the crystals \cite{supplementary}.

\begin{figure}[t]
\includegraphics[scale=.45]{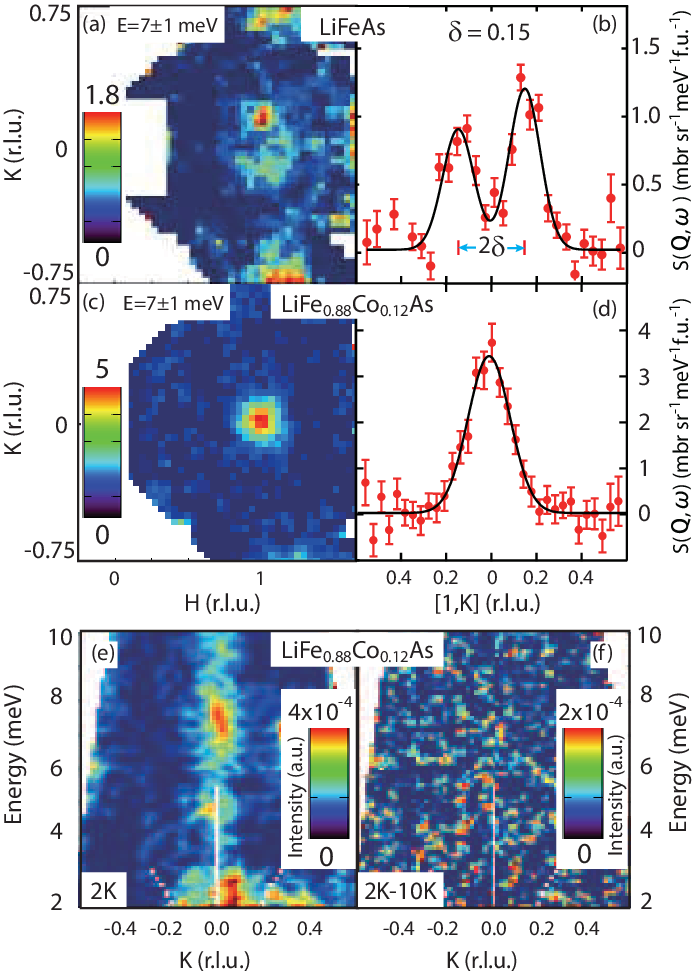}
\caption{(Color online) (a,c) Two-dimensional constant-energy images of spin excitations in the $[H,K]$ plane at $E=7\pm 1$ meV and 5 K for
LiFeAs and LiFe$_{0.88}$Co$_{0.12}$As, respectively.
The high scattering intensity near $(1,\pm 0.75)$ in (a) is due to phonons, but becomes less obvious
in (c) due to different intensity scale.
(b,d) Constant-energy cuts of spin excitations along the $[1,K]$ direction for  LiFeAs and LiFe$_{0.88}$Co$_{0.12}$As at $E=7\pm 1$ meV, respectively. The solid lines are fits to two Gaussians for LiFeAs and a single Gaussian for LiFe$_{0.88}$Co$_{0.12}$As \cite{supplementary}. The intensity is in absolute units by normalizing a vanadium standard. (e) Energy-momentum plots of spin excitations along the $[1,K]$ direction integrated from
$H=0.9$ to 1.1 for LiFe$_{0.88}$Co$_{0.12}$As. The commensurate spin excitations form a vertical rod of scattering centered at
${\bf Q}_{AF}=(1,0)$ point. (f) The temperature difference scattering between 2 K and 10 K.
}
 \end{figure}

We first compare
low energy spin excitations in pure LiFeAs ($T_c\approx 18$ K) and LiFe$_{0.88}$Co$_{0.12}$As ($T_c\approx 4$ K).
Figure 2(a) shows image of the $E=7\pm 1$ meV excitations near ${\bf Q}_{AF}$ for LiFeAs obtained on ARCS.
Consistent with earlier work \cite{Qureshi12,MWang12,Qureshi14}, the data reveals
clear transverse incommensurate spin excitations away from ${\bf Q}_{AF}$ as shown in the $[1,K]$ cut of Fig. 2(b).
The incommensurate peaks may arise from nesting of the
outer hole or inner hole Fermi surface to the electron Fermi surfaces, which
give slightly different incommensurability
$\delta_1$ and $\delta_2$, respectively, as seen in the experiment
[Fig. 1(c) and 1(e)] \cite{MWang12,Qureshi14}.
Figure 2(c) shows identical image of constant energy ($E=7\pm 1$ meV) excitations for LiFe$_{0.88}$Co$_{0.12}$As.  A constant energy cut
along the $[1,K]$ direction reveals that
spin excitations are well defined at the commensurate wave vector ${\bf Q}_{AF}$ [Fig. 2(d)].
Figure 2(e) shows the dispersion of commensurate spin excitations obtained on CNCS.  The rod like feature at ${\bf Q}_{AF}$ below 10 meV
confirms the commensurate nature of spin excitations in LiFe$_{0.88}$Co$_{0.12}$As.
To determine if weak superconductivity at $T_c=4$ K has an impact on low-energy spin excitations, we show in Fig. 2(f) temperature
difference plot between 2 K and 10 K.  The absence of the temperature difference scattering in Fig. 2(f) below and above $T_c$
suggests that the weak superconductivity
has negligible effect on the low-energy spin excitations.  Based on data in Fig. 2, we summarize
in Fig. 1(b) the Co-doping evolution of
the low-energy spin excitations in LiFe$_{1-x}$Co$_{x}$As.  Different from BaFe$_{2-x}$Ni$_x$As$_2$, where
the low-energy spin excitations becomes transversely incommensurate with increasing Ni-doping \cite{hqluo},
Co-doping in LiFeAs changes transversely incommensurate spin excitations to commensurate as shown in Fig. 1(b), Fig. 1(e) and 1(f).
The differences in the electron doping evolution of the low-energy spin excitations between
LiFe$_{1-x}$Co$_{x}$As and BaFe$_{2-x}$Ni$_x$As$_2$
 can be understood within the Fermi surface nesting
picture as due to the differences in Fermi surfaces of LiFeAs \cite{ZRYe,HMiao15} and BaFe$_2$As$_2$ \cite{myi}.
A unique feature of the Fermi surfaces in LiFeAs is the large $d_{xy}$ orbital hole pocket at $(1,1)$ [Fig. 1(c)] \cite{Borisenko}.
Upon Co-doping to introduce additional electrons to LiFeAs, the large $d_{xy}$ hole pocket
shrinks and results in a better nesting with the electron pocket at (0,1), while the small $d_{yz}/d_{xz}$
 hole pocket sinks below the Fermi level [Fig. 1(c) and 1(d)].  For LiFe$_{0.88}$Co$_{0.12}$As, the observed commensurate spin excitations
are consistent with this picture, and suggest that low-energy spin excitations are mostly driven from the $d_{xy}$ orbitals.
This is consistent with the random phase approximation (RPA) calculations using ARPES determined Fermi surfaces, where the low-energy
spin excitations for Co-doped LiFeAs
involve mostly the $d_{xy}$-$d_{xy}$ character (Fig. S3) \cite{supplementary}.
Similarly, spin excitations from the $d_{yz}$-$d_{yz}$ channel are considerably reduced with the suppression of superconductivity.

\begin{figure}[t]
\includegraphics[scale=.48]{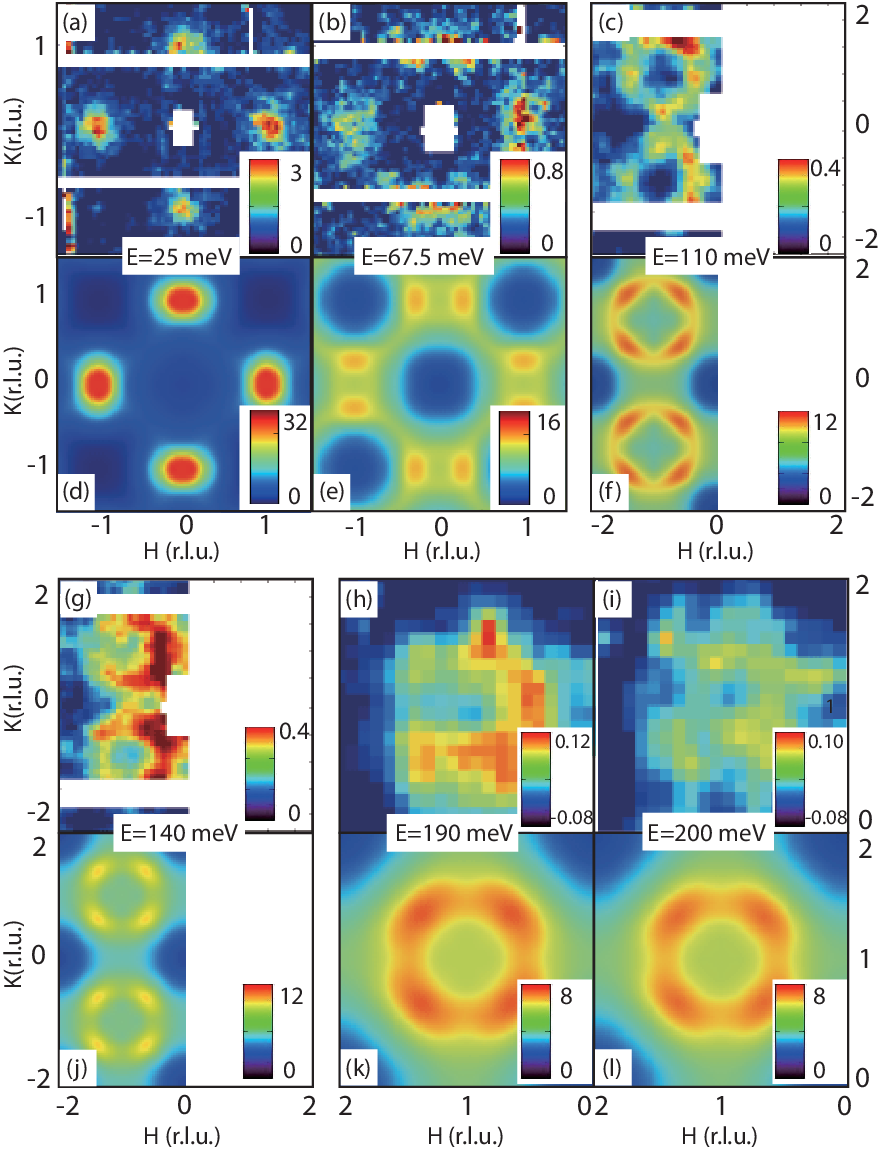}
\caption{(Color online) (a-c),(g-i) Constant-energy scattering images in the $[H,K]$ zones for LiFe$_{0.88}$Co$_{0.12}$As at energy transfers of $E=25\pm 5$ meV ($E_i= 80$ meV), $67.5\pm 7.5$ meV ($E_i= 250$ meV), $110\pm 10$ meV, $140\pm 10$ meV, $180\pm 10$ meV, and $200\pm 10$ meV ( $E_i=450$ meV). The scattering intensity is obtained after subtracting a radial background and has two-fold [(c),(g)] or fourfold symmetry [(h),(i)]. (d-f), (j-l) Corresponding constant-energy slices of dynamic magnetic structure factor $S({\bf Q},E)$ obtained from DFT+DMFT calculation. All data are taken at $T=5$ K.}
\end{figure}

Figure 3 summarizes the two-dimensional images of
spin excitations at different energies and their comparison with DFT+DMFT calculations for LiFe$_{0.88}$Co$_{0.12}$As.
Below $E=25$ meV, spin excitations occur at ${\bf Q}_{AF}=(1,0)$ and $(0,1)$ positions similar to spin waves
in NaFeAs [Fig. 3(a)] \cite{CLZ14}.  On increasing energy to $E=67.5\pm 7.5$ meV, spin excitations begin to
split vertically from $(1,0)$, again similar to spin waves of NaFeAs [Fig. 3(b)].
However, at energies above $E=100$ meV, spin excitations in LiFe$_{0.88}$Co$_{0.12}$As form rings of scattering centered around
$(\pm 1,\pm 1)$ which shrink slowly with increasing energy and persist up to $E=200$ meV [Figs. 3(c),
3(g), 3(h), and 3(i)].  This is significantly different from NaFeAs, where spin waves reach the band top near 100 meV \cite{CLZ14}.
Since high-energy spin excitations in LiFeAs behave similarly \cite{MWang11}, we conclude that spin excitations of
LiFe$_{1-x}$Co$_{x}$As have larger band width than that of NaFeAs \cite{CLZ14}, are similar to that
of BaFe$_{2-x}$Ni$_x$As$_2$ \cite{MSLiu,HQLuo13}.
Given the similar crystal structure and superconducting transition temperatures of LiFe$_{1-x}$Co$_{x}$As \cite{MJPitcher10} and
NaFe$_{1-x}$Co$_{x}$As \cite{Parker}, one would expect similar electron correlations and spin excitations band width in these two families of
materials \cite{ZPYin11,ZPYin14}.

\begin{figure}[t]
\includegraphics[scale=.5]{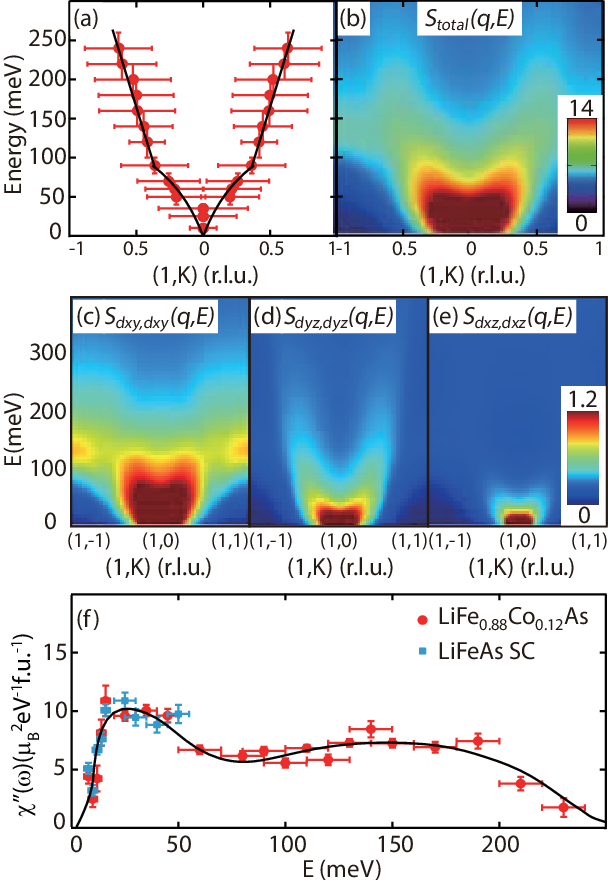}
\caption{(Color online) (a) The dispersion of spin excitations from time-of-flight neutron scattering data as seen in Fig. S2.
The points represent the peak positions
fitted with Gaussians. The errors in energy are the energy integration range and the {\bf Q}-errors come from the fitted peak width.
(b) The corresponding total dynamic spin susceptibility calculated by DFT+DMFT.
(c-e) The diagonal components of the dynamic magnetic structure factor $S_{xy,xy}({\bf q},E)$, $S_{yz,yz}({\bf q},E)$, and $S_{xz,xz}({\bf q},E)$ which originate from the $d_{xy}$, $d_{yz}$ and $d_{xz}$ orbitals, respectively.
(f) Energy dependence of the measured local dynamic spin susceptibility for LiFe$_{0.88}$Co$_{0.12}$As
and superconducting LiFeAs at $T=5$ K.
}
\end{figure}

To determine the spin excitation
dispersions of LiFe$_{0.88}$Co$_{0.12}$As, we made a series of cuts
on images of spin excitations in Fig. 3 along the $[1,K]$
direction at different energies (Fig. S2)
\cite{supplementary} and extracted the dispersion as shown in Fig. 4(a).  Compared with dispersions of spin waves in NaFeAs \cite{CLZ14} and
spin excitations in BaFe$_{2-x}$Ni$_x$As$_2$ \cite{MSLiu,HQLuo13}, dispersion of LiFe$_{0.88}$Co$_{0.12}$As has distinctive features around 100 meV [Fig. 4(a)].  Figure 4(b) shows the DFT+DMFT calculated total dynamic spin dynamic susceptibility, which
reveals clear two component structure similar to spin excitations in Fig. 4(a).  Figures 4(c), 4(d), and 4(e) are
dynamic spin susceptibility corresponding to
the $d_{xy}$-$d_{xy}$, $d_{xz}$-$d_{xz}$, and $d_{yz}$-$d_{yz}$ intra-orbital scattering channels between the hole and electron Fermi surfaces, respectively. Along the $(1,K)$ direction, spin excitations from the $d_{xy}$ orbital
reach zone boundary around $E=130$ meV [Fig. 4(c)], while excitations from the $d_{yz}$ orbital extend to energies well above
$E=200$ meV  [along the $(K,1)$ direction, it would be the $d_{xz}$ orbital component due to the four fold symmetry].
The similarities in Figures 4(a) and 4(b) strongly suggest
that the upper and lower branches of the observed spin excitations have different orbital origins.
In Figure 4(f), we compare the estimated local dynamic spin susceptibility for LiFe$_{0.88}$Co$_{0.12}$As and LiFeAs at $T=5$ K.
The total fluctuating moment of LiFe$_{0.88}$Co$_{0.12}$As is $\left\langle {\bf m}^2\right\rangle=1.5\pm 0.3\ \mu_B^2$/Fe.
This is similar to superconducting LiFeAs \cite{Qureshi12,MWang12,Qureshi14}, but is somewhat smaller than
those of NaFeAs ($\left\langle {\bf m}^2\right\rangle\approx 3.2\ \mu_B^2$/Fe) \cite{CLZ14}
and BaFe$_2$As$_2$ ($\left\langle {\bf m}^2\right\rangle\approx 3.6\ \mu_B^2$/Fe) \cite{Harriger,Harriger12}.
This means that the total fluctuating
moments for LiFeAs family of materials are smaller than those of NaFeAs and BaFe$_2$As$_2$ iron pnictides.

In iron pnictide such as BaFe$_2$As$_2$, spin wave dispersions can be well described by an anisotropic
Heisenberg Hamiltonian \cite{Harriger}. However, the two branch feature of the spin excitation dispersion in LiFe$_{0.88}$Co$_{0.12}$As
clearly cannot be satisfactorily fitted by this anisotropic Heisenberg model. Our neutron scattering experiments and DFT+DMFT calculations suggest that orbital selective quasiparticle excitations may account for the energy and wave vector dependence of spin excitations in LiFe$_{0.88}$Co$_{0.12}$As. This indicates that the superexchange spin interactions are different for different orbitals.

It is well known that electronic correlations
in iron pnictides depend sensitively on the Fe pnictogen distance owing to the kinetic frustration mechanism of the
Fe $3d$ electrons, and are strongly enhanced with increasing
Fe-pnictogen distance \cite{ZPYin11,ZPYin14,YKKim}. Together with the large Hund's rule coupling and strong on-site Coulomb repulsion, the kinetic frustration mechanism also gives rise to the strong orbital differentiation of the electronic correlation strength \cite{ZPYin11,Georges}.
Orbital selective electronic correlation has been found in FeTe$_{1-x}$Se$_x$, where the effective mass of bands dominated by the $d_{xy}$ orbital character decreases with increasing selenium as compared to the $d_{xz}/d_{yz}$ bands \cite{ZKLiu}. In the case of LiFeAs, charge transfer from the $d_{xy}$ to $d_{xz}/d_{yz}$ orbitals can account for the Fermi surface topology of LiFeAs as the consequence of orbital dependent
band renormalization \cite{ZPYin11,GLee}. As shown in Fig. S6 \cite{supplementary}, the increased pnictogen height in LiFeAs compared with NaFeAs narrows the electronic bandwidth of the $d_{xy}$ orbital, which in turn transfers electrons from the $d_{xy}$ to the $d_{xz}/d_{yz}$ bands. The observed Co-doping dependence of low-energy spin excitations results from the $d_{xy}$-$d_{xy}$ orbital dependent Fermi surface nesting. The narrow electronic bandwidth of the $d_{xy}$ also leads to narrow bandwidth of spin excitations, and weak effective magnetic exchange coupling.

Since the $d_{xy}$ orbital dominated Fermi surface nesting becomes better for LiFe$_{0.88}$Co$_{0.12}$As, low-energy spin excitations become
commensurate with enhanced spectral weight compared to incommensurate spin excitations in LiFeAs [Fig. 1(b) and 1(d)].  This is consistent with
NMR measurements \cite{YMDai} and RPA/DFT+DMFT calculations (Figs. S3 and S4) \cite{supplementary}.
The observed non-Fermi liquid behavior near $x=0.12$ is then due to vanishing Fermi surface pocket
associated with $d_{yz}/d_{xz}$ orbitals
as the Lifshitz transition is approached from the underdoped side \cite{SDDas}.
In principle, an increased spin-fluctuation spectral weight should provide a larger electron pairing strength, and thus higher $T_c$ within the spin-fluctuation mediated
superconductivity scenario \cite{Scalapino}. However, since
Co-doping to LiFeAs also induces large incoherent electron
scattering \cite{ZRYe} and narrows the magnetic bandwidth in the $d_{xy}$ orbital [Fig. 4(c)], superconductivity
associated with the $d_{xy}$ orbital may be prohibited due to reduced effective magnetic exchange coupling associated with the
$d_{xy}$ orbitals \cite{MWang13}.
  Similarly,
in spite of the large magnetic bandwidth associated with the
$d_{xz}/d_{yz}$ orbitals, the poor Fermi surface nesting of these orbitals suppresses
low energy spin excitations, which is also bad for superconductivity \cite{MWang13}.
Therefore, superconductivity in iron pnictides can only occur with appropriate orbital selective
low-energy spin excitations coupled with reasonable large magnetic exchange coupling.

The neutron scattering work at Rice is supported by the
U.S. DOE, BES DE-SC0012311 (P.D.).  The computational work at Rice, ORNL, and Rutgers
is supported by NSF DMR-1436006 (P.D.),
DMR-1308603 (T.M.), DMR-1405303 (K.H.), and DMR-1308141 (Z.P.Y. and G.K.).
The materials effort at Rice is also supported by
the Robert A. Welch Foundation Grant Nos. C-1839 (P.D.).
The research at SNS was
sponsored by the Scientific User Facilities Division, BES, U.S. DOE. The research
used resources of the Oak Ridge Leadership Computing Facility at ORNL, which is supported by
U.S. DOE under Contract No. DE-AC05-00OR22725. The work at IOP, CAS is supported by
NSFC and MOST of China through research projects.

\pagebreak
\widetext
\begin{center}
\textbf{\large Supplementary Information: }
\end{center}
\setcounter{equation}{0}
\setcounter{figure}{0}
\setcounter{table}{0}
\setcounter{page}{1}
\makeatletter
\renewcommand{\theequation}{S\arabic{equation}}
\renewcommand{\thefigure}{\arabic{figure}}
\renewcommand{\bibnumfmt}[1]{[S#1]}
\renewcommand{\citenumfont}[1]{S#1}

\section{Sample preparation}
LiFe$_{1-x}$Co$_x$As single crystals were grown with self-flux method. The basic sample characterization was
described in the previous papers \cite{sref1,sref2}. Samples used in this report were grown with isotope $^7$Li to reduce the
neutron absorption and wrapped by Aluminum foil with Hydrogen-free glue to avoid exposure to air and
humidity. The sample growth work was carried out at Beijing National Laboratory for Condensed Matter
Physics, Institute of Physics, Chinese Academy of Science and at Rice University.
\section{Background subtraction and data analysis}
In a typical time-of-flight experiment, the raw inelastic neutron scattering data at certain energy are shown
in Fig. S1(a). In order to obtain the background, we masked the signal area, for instance, the white square in
Fig. S1(b). Assuming the background in our time-of-flight data is radially symmetric, we integrated the
remaining intensity and fitted it with a polynomial function of $\mid$Q$\mid$ to the second order as shown in Fig. S1(c).
Then we used this fitted polynomial function as background and subtracted it from the raw neutron
scattering data. The final subtracted data was shown in Fig. S1(d) and Fig. S1(e).

In Fig. S2, we show a series of typical subtracted constant-energy cuts from 10 meV to 220 meV. These
cuts were fitted with one or two Gaussian functions. We show the fitted result in previous Fig. 4(a). It is
worth noting that the x-errors in Fig. 4(a) are the fitted peak width.
\section{RPA calculation}
In Fig. S3, we demonstrate the existence of two transversely incommensurate peaks in the dynamic spin
susceptibility with differentiated orbital character for LiFeAs. Electron-doping in LiFe$_{0.88}$Co$_{0.12}$As is
introduced by rigid band shift. The starting point for our calculation is an effective 10-orbital tight-binding
Hamiltonian derived from ARPES and symmetry considerations \cite{sref4} and previously discussed in Ref. \cite{sref5}.
We calculate the noninteracting bare susceptibility along high-symmetry cuts, considering lowest-order
scattering processes as described previously \cite{sref5,sref6}:
\begin{equation}
\chi ^0_{l_1l_2l_3l_4}(q,\omega) =-\frac{1}{N} \sum _{k,\mu\nu} \frac{a^{l_4}_{\mu}(k) a^{l_2,*}_{\mu}(k) a^{l_1}_{\nu}(k+q) a^{l_3,*}_{\nu}(k+q)}{\omega+E_{\mu}(k)-E_{\nu}(k+q)+i\delta} (f[E_{\mu}(k),kT]-f[E_{\nu}(k+q),kT])
\end{equation}
with N = 2 the number of iron sites per unit cell, band indices $\mu$ and $\nu$, orbital indices l . The matrix
elements are represented by the orbital projection of the Bloch state, $a^l_\mu$=$<l\mid\mu$k$>$ and f$[$E,$kT]$ is the
Fermi function at temperature T . We use T= 100 K , a small parameter $\delta$=0.005 to enforce analyticity,
and sum over a k-space mesh of 120$\times$120$\times$8 points over the 3D Brillouin zone, which we find to be
sufficiently dense to accurately describe the susceptibility everywhere in reciprocal space.

RPA interactions are applied in a manifestly spin-rotationally invariant form, using the intra-orbital
Coulomb repulsion U = 0.6 eV and Hund¡¯s coupling J = 0.15 , and inter-orbital Coulomb interaction
U'=0.3 and pair-hopping J'= 0.15, with the same interaction strengths applied identically over all ten d-orbitals. These values for the interactions are just below their maximum values, as determined when the
susceptibility diverges at the antiferromagnetic wave vector Q in the doped compound. Because the intraorbital scattering is dominant in the total RPA susceptibility, we focus on intra-orbital components of the
bare susceptibility in the low energy limit, $\chi^0$(q,$\omega$= 10 meV). To enable direct comparison to neutron
scattering, where the orthorhombic structure factor forbids scattering at even L, we fix q$_z$=$\frac{\pi}{c}$ (L=1).

We find the scattering peaks in the parent compound ($<$n$>$=12.00 ) at $\Delta$Q$\thickapprox$0.18 and 0.3, quite consistent
with the incommensurabilities suggested by Fermi surface nesting shown in Fig. 1(c)-(d). After the
application of RPA interactions, tracing only over the d$_{xy}$ (d$_{xz}$) orbitals selectively amplifies the narrower
(wider) peaks; in other words, considering the system with and without interactions, the results are
qualitatively identical. In the electron-doped case ($<$n$>$=12.24 ), the single obvious feature is the
commensurate peak, which has predominantly d$_{xy}$ character.
\section{DFT+DMFT calculation}
Density functional theory (DFT) calculations were done using the full-potential linear augmented plane
wave method implemented in Wien2K\cite{sref7} in conjunction with a generalized gradient approximation\cite{sref8} of the
exchange correlation functional. To take into account strong correlation effect, we further carried out first principles calculations using a combination of density functional theory and dynamical mean field theory
(DFT+DMFT)\cite{sref9} which was implemented on top of Wien2K and documented in Ref. \cite{sref10}. In the
DFT+DMFT calculations, the electronic charge was computed self-consistently on DFT+DMFT density
matrix. The quantum impurity problem was solved by the continuous time quantum Monte Carlo (CTQMC)
method\cite{sref11,sref12}, using Slater form of the Coulomb repulsion in its fully rotational invariant form. Consistent
with previous publication\cite{sref13,sref14}, we used a Hubbard U=5.0 eV and Hund's rule coupling J=0.8 eV, and
experimentally determined crystal structure for LiFeAs\cite{sref15}, including the internal positions of the atoms.
The Co-doping is simulated using virtual crystal approximation (VCA) in the DFT+DMFT calculation.
The bare susceptibility was computed using the fully self-consistent DFT+DMFT lattice Green's function
and the spin susceptibility was computed using the Bethe-Salpeter equation which takes into account two particle vertex correction. Here the two-particle (particle-hole) irreducible vertex is local within DMFT and
it is equal to the impurity vertex, which can be obtained from the solution of the quantum impurity model
using CTQMC. Further computational details on spin susceptibility are available in Ref.\cite{sref14}.

\begin{figure}[t]
\includegraphics[scale=.95]{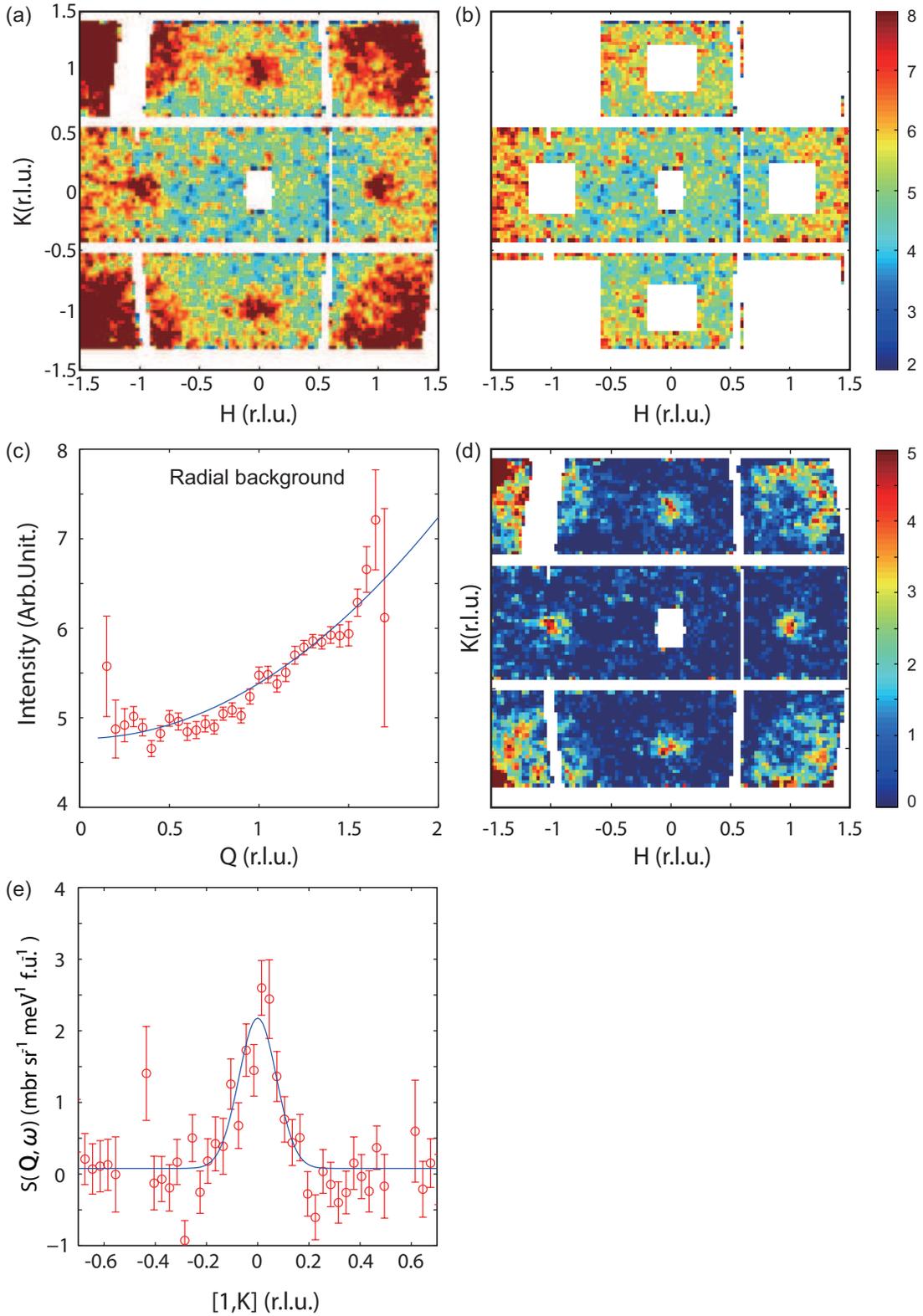}
\caption{(Color online) (a).Constant energy map of the raw data at E=$[$6,8$]$ meV in LiFe$_{0.88}$Co$_{0.12}$As. (b)
The background intensity in which all the phonon and magnon signals are masked. (c) We get the radially
symmetric background from (b) and fit it with a parabolic function. (d) The constant energy map after
background subtraction. (e) A cut along $[$1,H$]$ direction in panel (d).}
\end{figure}

\begin{figure}[t]
\includegraphics[scale=.95]{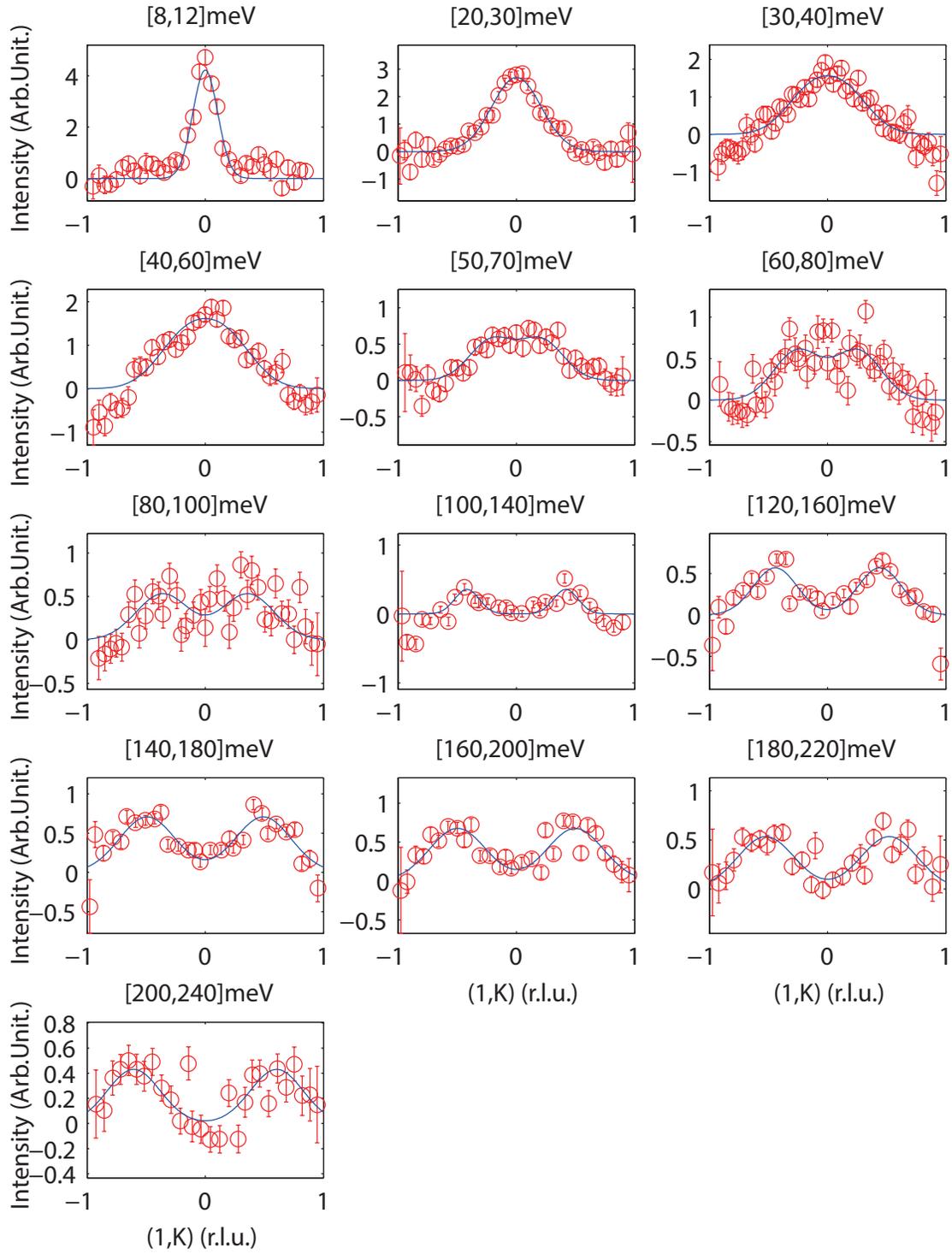}
\caption{(Color online)
Energy dependence of the 1-d cut of the spin excitations along $[$1, K$]$ direction
in LiFe$_{0.88}$Co$_{0.12}$As. The peaks are fitted with Gaussian functions and the fitted results are shown in Fig. 4(a).
The energy ranges shown here are the E-errors and the peak widths are q-errors Fig. 4(a).
}
\end{figure}

\begin{figure}[t]
\includegraphics[scale=.95]{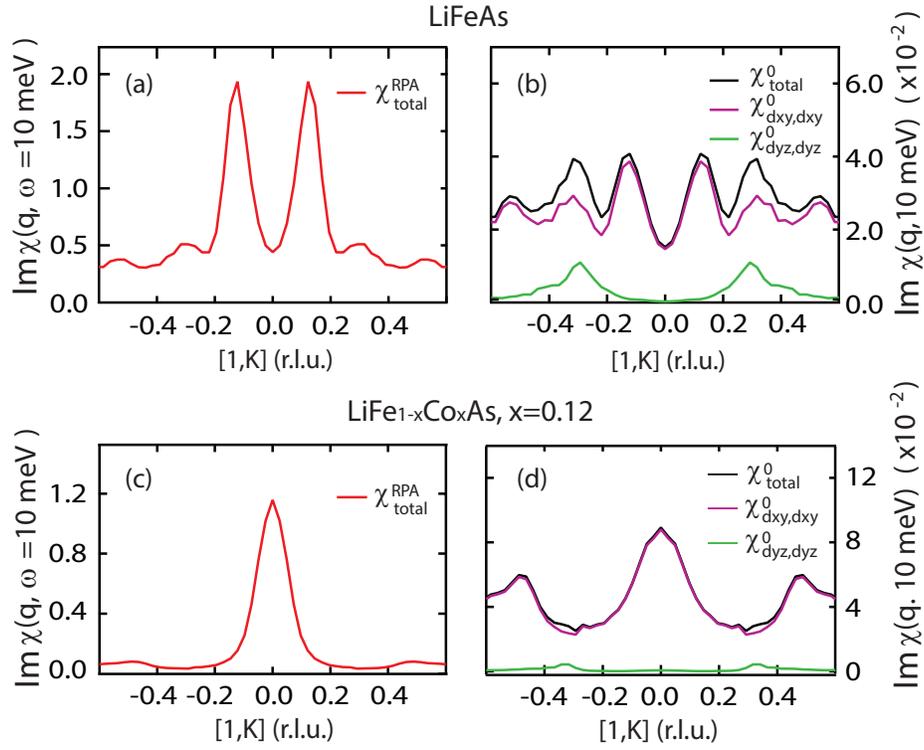}
\caption{(Color online) LDA+RPA calculation of dynamic spin susceptibility in LiFeAs and
LiFe$_{0.88}$Co$_{0.12}$As. (a),(c) Wave vector dependence of the imaginary part of total RPA susceptibility at 10
meV for LiFeAs and LiFe$_{0.88}$Co$_{0.12}$As. Note that the peak position changes from incommensurate wave
vector in LiFeAs to commensurate in 12$\%$ electron doped compound, consistent with the our experimental
result (Fig.1 (b)). (b),(d) The calculated bare susceptibility of LiFeAs and LiFe$_{0.88}$Co$_{0.12}$As. The black curve
represents the total bare susceptibility while the purple (green) one shows the intra-orbital component of d$_{xy}$
(d$_{yz}$) orbital. Note that the main peaks observed in LiFeAs mainly come from
$\chi_{xy,xy}$ scattering channel while
the
$\chi_{yz,yz}$, (or $\chi_{xz,xz}$ due to the existence of four-fold symmetry) component almost vanishes when T$_c$ is
suppressed in LiFe$_{0.88}$Co$_{0.12}$As.}
\end{figure}

\begin{figure}[t]
\includegraphics[scale=.95]{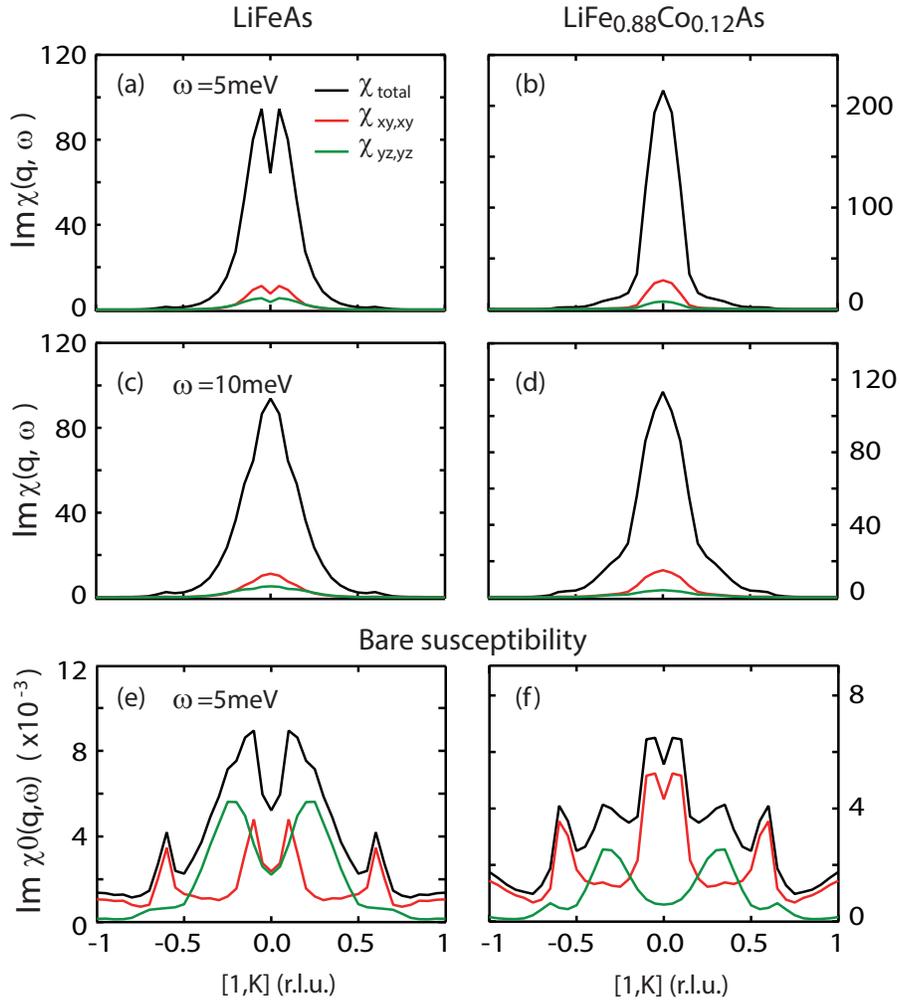}
\caption{(Color online)
The dynamic spin susceptibility from DFT+DMFT calculation for LiFeAs and
LiFe$_{0.88}$Co$_{0.12}$As. (a),(b) The imaginary part of spin susceptibility at 5 meV for LiFeAs and LiFe$_{0.88}$Co$_{0.12}$As,
respectively. The black curves represent the total susceptibility and the red (green) ones are the
$\chi_{xy,xy}$( $\chi_{yz,yz}$)
components. (c),(d) The corresponding total and orbital components of spin susceptibility at 10 meV for
LiFeAs and LiFe$_{0.88}$Co$_{0.12}$As. (e), (f) are bare susceptibility for LiFeAs and LiFe$_{0.88}$Co$_{0.12}$As, respectively, at
5 meV. The doping dependence of the incommensurability is similar to the experimental result and
LDA+RPA calculation. Note that in 12$\%$ Co doped compound, the
$\chi_{xy,xy}$ component is actually enhanced
while
$\chi_{yz,yz}$, component is suppressed. The overall intensity of total susceptibility within this energy range is
also enhanced a little bit, in sharp contrast to great suppression the superconducting temperature (T$_c$) in this
compound.
}
\end{figure}

\begin{figure}[t]
\includegraphics[scale=.95]{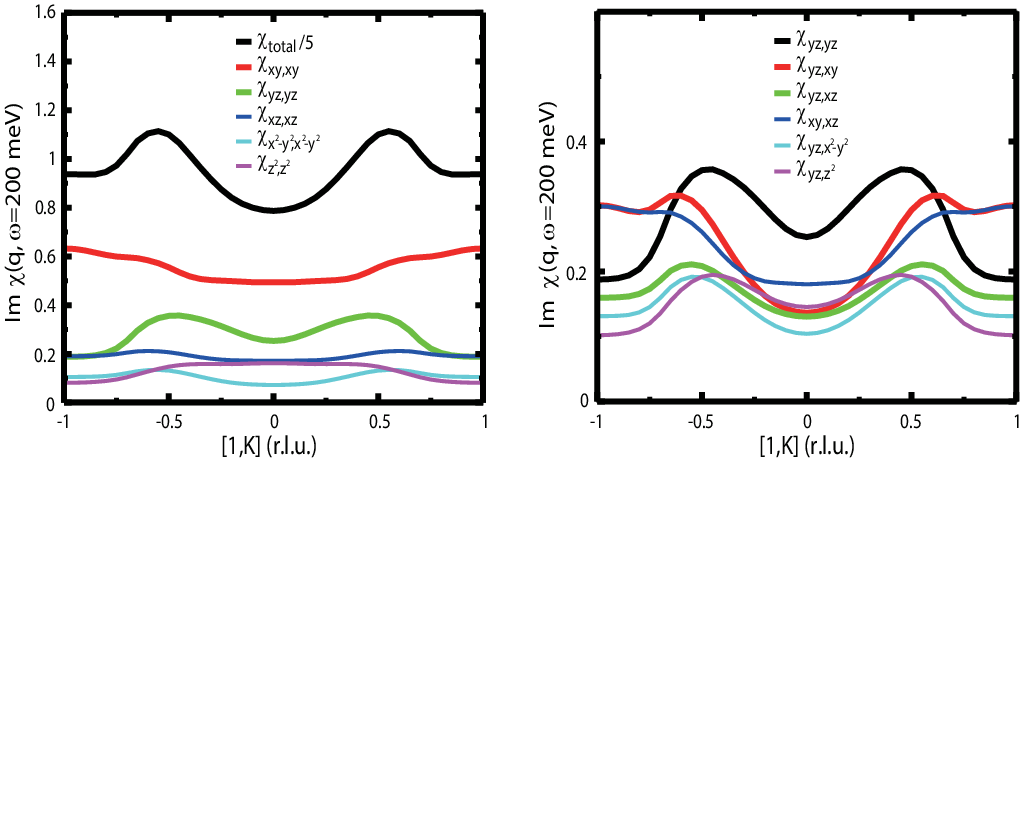}
\caption{(Color online) (a) The intra-orbital components of dynamic susceptibility at 200 meV
(mainly upper branch) for LiFe$_{0.88}$Co$_{0.12}$As. It is clearly seen that the peaks in the total susceptibility mainly
come from
$\chi_{yz,yz}$ intra-orbital part. (b) Comparison between intra-orbital and inter-oribtal components which
are associated with d$_{yz}$ orbital. The intra-orbital
$\chi_{yz,yz}$, component is apparently dominant while the interorbital channels contribute a small part in the total spin susceptibility at 200 meV.}
\end{figure}

\begin{figure}[t]
\includegraphics[scale=.95]{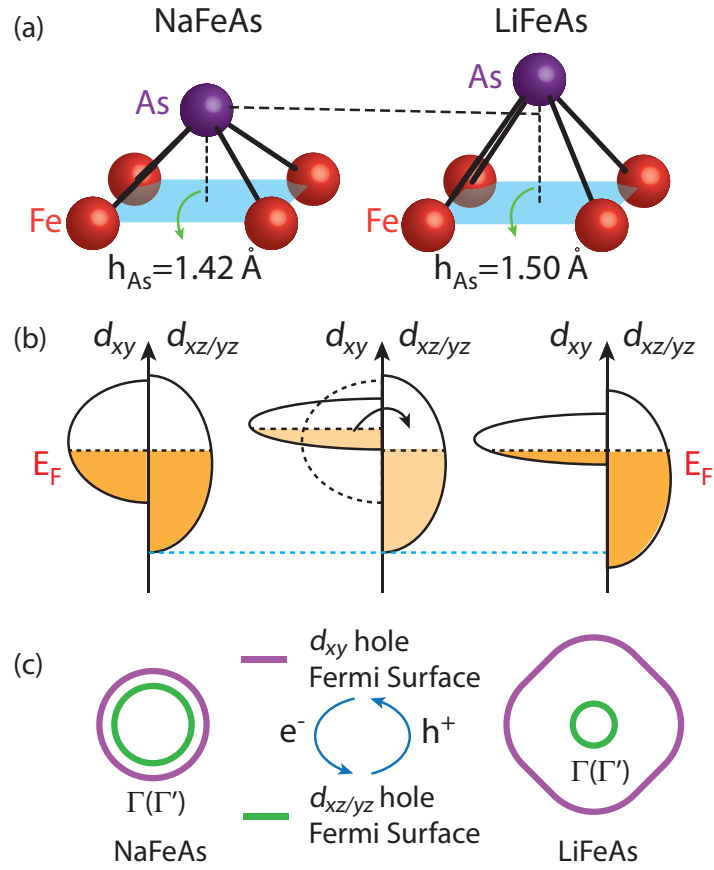}
\caption{(Color online) (a) The structure of the Fe-As layer in NaFeAs and LiFeAs. The anion height
in LiFeAs $~$ 1.5 ${\AA}$(Ref. \cite{sref2} ) is larger than that in NaFeAs ~1.42 ${\AA}$ (Ref. \cite{sref4} ) . (b) Illustration of the orbital
dependent band renormalization and subsequent charge transfer from d$_{xy}$ to d$_{xz/yz}$ orbitals. The
renormalization is overstated to emphasize the orbital dependent effect. The situation in reality might be
even more complicated due to the strong hybridization with As p orbitals \cite{sref6}. (c) Comparison of the Fermi
Surface in NaFeAs and LiFeAs. The orbital dependent band renormalization (panel (b)) drives electrons
(holes) from d$_{xy}$ (d$_{xz/yz}$) to d$_{xz/yz}$ (d$_{xy}$), resulting in an enlarged outer d$_{xy}$ hole FS and a reduced inner d$_{xz/yz}$
pockets in LiFeAs.}
\end{figure}

\end{document}